\documentclass{article}
\usepackage{amsmath}
\usepackage{amscd}
\usepackage{amsthm}
\usepackage{amssymb} \usepackage{latexsym}
\usepackage{eufrak}
\usepackage{euscript}
\usepackage{epsfig}
\usepackage{graphics}
\usepackage{array} 
\usepackage{enumerate}
 \usepackage{boxedminipage}
\usepackage{pictexwd,dcpic}
\usepackage{ulem}

\usepackage{color}

\usepackage{hyperref}

\newcommand{\bel}[1]{\begin{equation}\label{#1}}

\newcommand{\be}{\begin{equation}}

\newcommand{\ba}{\begin{eqnarray}}
\newcommand{\ea}{\end{eqnarray}}
\newcommand{\rf}[1]{(\ref{#1})}
\newcommand{\bi}{\bibitem}
\newcommand{\qe}{\end{equation}}

\newcommand{\R}{{\mathbb R}}

\newcommand{\Cat}[1]{\mathbf{#1}}
\newcommand{\Hom}{\mathrm{Hom}}

\newtheorem{thesis}{Thesis}
\newcommand{\btl}[1]{\begin{thesis}\label{#1}}
\newcommand{\et}{\end{thesis}}

\theoremstyle{theorem}

\theoremstyle{corollary}

\theoremstyle{lemma}
\newtheorem{lemma}{Lemma}[section]
\theoremstyle{definition}
\newtheorem{defi}{Definition}[section]
\theoremstyle{remark}
\newtheorem*{pf}{Proof}
\theoremstyle{remark}

\title{A mathematical model for cognitive structures and processes underlying innovations}
\author{J\"urgen Jost and Massimo Warglien}
\begin{document}
\maketitle
\begin{abstract}
    Business innovations are often arising from new combinations of pre-existing systems into new ones, coherently assembling  features from their sources.  We propose an abstract mathematical concept from category theory, the presheaf, to efficiently represent such coherent feature combinations. Moreover, operations on presheaves will allow us to formally describe and analyze business innovations that arise from novel mergings of systems from different domains. Equipped with such tools we provide an example by analyzing a  successful case of such type of recombinant innovation, the digital hub concept proposed by Steve Jobs. The example shows how our framework can be used to bring formal rigor while preserving a fundamentally qualitative reasoning style.
\end{abstract}


\section{Introduction} 
Innovations often arise as novel combinations of existing structures or systems. Since at least Adam Smith  “combining together the powers of the most opposite and distant objects” has been recognized as a distinctive capability underlying innovation \cite{Sm}. For Schumpeter\cite[p.100f]{Sch1}, new combinations (new products, new production methods, new markets, new ressources, or new market organisations) are the key to entrepreneurial success; he writes  “innovation combines factors in a new way, or (…) it consists in carrying out New Combinations.” \cite{Sch}] . The recent literature on innovation has repeatedly emphasized notions such as ‘resource recombination \cite{GR}], ‘exploration as recombination’ \cite{YS}. The process through which different conceptual structures are recombined has been also the object of important research in cognitive science \cite{FT} and studies of combinatorial creativity \cite{Boden}. However, a formal representation of how different structures or systems can be combined to produce innovation is still missing. In our work, we want to provide a systematic mathematical framework for understanding the representation and organization, the modification and reorganization, and the combination of systems in a coherent fashion. The mathematical framework we use is based on category theory and preheaves – an approach that allows to combine formal rigor with an essentially qualitative reasoning about structural relationships in strategic thinking.\\

We shall thus first develop the mathematical framework. That framework is
taken from category theory, and the notion of a presheaf is most basic for our
purposes. For the mathematical theory and further references, see \cite{J1}.\\
We then provide a detailed case study, Steven Jobs' well-known digital hub concept\cite{Jo}.The
abstract structure underlying this hub 
can be coherently analyzed in our framework. By seeing the abstract pattern,
strategists should be enabled to transfer it to other situations. \\

The basic idea is the following. A successful business strategy has to consist
of a coherent or fit collection of the values of various features. A well-known example comes from the low-cost flight industry, where the combination of coherent activities such as limited passenger service, lean ground services, frequent flies, point-to-point routes, high aircraft utilization and low prices sustained the innovative strategy of Southern Airlines \cite{Po2}.  Arbitrary combinations of such feature values will not
constitute a viable strategy. Only a small subset of features' combinations is to be considered in an efficient representation. In particular, the more features have to be
taken into account, the more constraints there will be in general for the
combinations of their values. While at first glance, this may seem a problem,
we can turn it to our advantage. When we know some feature values, we can
already infer the ranges of others, and we no longer need to bother about
arbitrary values. Formally, we represent this by a presheaf and its
sections. Only sections that correspond to feasible combinations are
permitted.
Moreover, we systematically analyze how business strategy involving different
sets of features can be merged or amalgamated to produce true innovations. We
exemplify the abstract formalism with Steve Jobs' conception of a digital hub concept
that enabled Apple to become a dominant player in the digital world.

\section{Presheaves and the organization of structures}
In the most trivial way, a system consists of elements and relations. If we
have $N$ elements, then there are $N^2$ possible pairwise
relations,\footnote{If we only consider symmetric relations, that is, $A$
  standing in a relation with $B$ is the same as $B$ standing in a relation
  with $A$, and if we exclude relations of an element to itself, then there
  are still $\frac{N(N-1)}{2}\approx \frac{N}{2}$ possibilities.}  $N^3$
triple ones, and so on. Thus, if such pairwise relation can be present or
absent, there are $2^{N^2}$ possibilities. When for instance $N=100$, then
$N^2=10.000$, and $2^{10.000}$ is astronomically larger than the number of
particles in the universe. One speaks of a combinatorial explosion here, and
this does not seem to be an efficient road towards encoding the structure of a
system.\\
Of course, we are only interested in particular types of systems, and not in all
possible ones. But if we want to explore novel ones, how can we restrict those
that are worth exploring? Just consider those that are close to the already
known ones? But that would not account for true novelties that may be quite
different from any previously known structure.\\
\\
In order to bring content into the picture, we want to describe a system in terms 
of features or observables and their values. To resort to another familiar example, in the wine industry relevant features are price, wine complexity, aging quality, vineyard prestige and legacy...These features are not independent from each other and premium wines differ from budget wines in the way they combine them.  For example, high price will be consistent with good aging quality and wine complexity, and characterize premium wine strategy as compared with budget wines that will combine lower values of such features \cite{KM}.

We therefore want to introduce a
formal framework that can efficiently account for and exploit such relations
between feature values. \\

The framework that we utilize is that of
presheaves. As this is an advanced mathematical theory,  a reader that is mainly interested in the applications to innovations may skip
that part on a first reading and only return to it when trying to understand
the formalism underlying those applications. 

\subsection{Categories and presheaves}
Category theory is an abstract theory of structures. Objects in a category
are completely characterized by their relations with other objects and 
do not possess intrinsic properties beyond that. Here are  the
formal definitions.

\begin{defi}A  \textit{category} $\mathbf{C}$\index{category} consists of \textit{objects}\index{object} $A,B,C,\dots $ and \textit{ arrows}\index{arrow} or \textit{morphisms}\index{morphism}
$ f:A\to B$ between objects. They have to satisfy some properties:
\begin{itemize}
\item  Arrows can be composed: 
$f:A\to B$ and $g: B\to C$ generate the  arrow
\begin{equation}\label{cat1-2}
g\circ f:A \to C.
\end{equation}
This composition is \textit{associative},\index{associative}
\begin{equation}\label{cat1-3}
h\circ (g\circ f)=(h\circ g)\circ f
\end{equation}
for $f:A\to B, g:B\to C, h:C\to D$.
\item For each object $A$, we have the \textit{identity
arrow} \index{identity arrow}
\begin{equation}\label{cat1-4}
1_A:A\to A
\end{equation}
with 
\begin{equation}\label{cat1-5}
f\circ 1_A=f=1_B\circ f
\end{equation}
for all $f:A\to B$
\end{itemize}
\end{defi}
The presence of the 
identity arrow is implicitly
understood and therefore often not explicitly mentioned. Therefore, when we 
list the arrows in a category, we often only mention the others. \\

A given category need not have any morphisms beyond the identity morphisms,
that is, it could consist of objects with no morphisms whatsoever between
different objects, and even no other morphisms of any object to itself. In
fact, the simplest category $\mathbf{0}$ does not contain any objects or
morphisms at all. And the  category  $\mathbf{1}$ 
 has  only a single object with its identity morphism.  But a category could
also consist of a single object with several morphisms. The category
$\mathbf{2}$ has two objects, $1,2$, with the only non-trivial arrow $1\to 2$.
Such simple categories
are often used as indexing categories, in order to specify particular patterns
in other categories. Before coming to that, however, it will be useful to
develop some general principles and constructions.\\

From a category $\Cat{C}$, we can easily and naturally construct other
categories. For instance, we can take the morphisms of $\Cat{C}$ as the
objects of a new category $\Cat{C'}$. The morphisms $F$ of $\Cat{C'}$ then have to
relate the morphisms of the original category $\Cat{C}$, that is, for a
morphism $f:A\to B$, both $A$ and $B$ and the relation between them needs to
be transformed. That is achieved by a commutative diagram.
\begin{defi}A \textit{commutative} or \textit{commuting  diagram} is a
  relation of the form 
\begin{equation}\label{cat1-5a}
\begin{CD}
   A @>{f}>> B    \\
    @V{F}VV    @VV{G}V     \\
    A' @>{g}>>B'
  \end{CD}
\end{equation} 
with the relation
\begin{equation*}
  G\circ f=g\circ F,
\end{equation*}
thaat is, the result remains the same regardless of the sequence of arrows that we use
to get from the upper left to the lower right. 
\end{defi}
So, for a morphism $F:f\to g$ in $\Cat{C'}$, where $f:A\to B$ and $g:A'\to
B'$ are morphisms in $\Cat{C}$, we require that
\begin{equation}\label{cat1-5b}
\begin{CD}
   A @>{f}>> B    \\
    @V{F}VV    @VV{F}V     \\
    A'=F(A) @>{g}>>B'=F(B)
  \end{CD}
\end{equation}
commutes. Thus $g=F(f)$ transforms $F(A)$ into $F(B)$ such that the relation
between $A$ and $B$ established by $f$ is preserved.\\

A basic category is $\Cat{Sets}$ whose objects are sets.\footnote{We do not enter here into the foundational
  discussion of what should be accepted as a set. In category theory, one
  usually assumes a universe $\mathcal{U}$ of sets that enjoys certain
  properties. For instance, if $U,V\in \mathcal{U}$, then also all their
  subsets, the sets of
  unordered or ordered pairs of members of $U$ and $V$ are in $\mathcal{U}$,
  and so are their Cartesian product and union and the power set of $U$. Also,
  $\mathcal{U}$ should contain the set of ordinal numbers, to make it
  sufficiently rich. For more general discussions, one may invoke the
  Zermelo-Frankel or G\"odel-Bernays axioms of set theory. Of course, one
  wants to avoid logical paradoxes, like the set of all sets that do not
  contain themselves as elements.} There are two important possibilities for
the choice of morphisms.
\begin{enumerate}
\item  Subset relation: for two sets $U,V$, $U\to V$ means
  $U\subset V$.
  \item $f:V\to U$ is an arbitrary map between sets, without any further
    restriction. (Restrictions may arise when we impose additional strucutre,
    for instance consider the category of topological spacws where morphisms
    are \textit{continuous} maps.)
\end{enumerate}
This category, with either type of morphisms,  will occur repeatedly as some kind of
background category.\\

The characterization of objects in terms of relations naturally leads to
\begin{defi}\label{def:Category2}The objects $A$ and $B$ are
  \textit{isomorphic}\index{isomorphic} if there are 
morphisms $f:A\to B$ and $g:B\to A$ with
\begin{equation}\label{cat1-6}
g\circ f=1_A\quad \text{ and } f\circ g=1_B.
\end{equation}
\end{defi}

But the principle is much more general than that. 
\begin{defi}\label{def:Category3}For objects $C,D$ of a category $\Cat{C}$,
  \begin{equation}\label{presh1}
\Hom_\mathbf{C}(D,C)
\end{equation}
is the collection of all morphisms $D\to C$. 
  \end{defi}
In order to relate this to set theory, we require a {\bf Convention:}
For all  categories in this chapter, denoted by $\Cat{C}$
  or other symbols, 
$\Hom_\mathbf{C}(D,C)$ is a set,  called a \textit{$Hom$-set}.

Thus
\begin{equation}
  \label{presh2}
  D\mapsto \Hom_\mathbf{C}(D,C)
\end{equation}
maps each object $D$ of $\Cat{C}$ to a set, and a morphism $f:D\to E$ leads
to the map
\begin{eqnarray}
  \nonumber
  \Hom_\mathbf{C}(E,C) &\to& \Hom_\mathbf{C}(D,C)\\
  \label{presh3}
  (h:E\to C) &\mapsto& (h\circ f;D\to C)
\end{eqnarray}
There are two aspects that we can observe and that we shall generalize
subsequently:
\begin{enumerate}
\item An object $D$ and a  morphism $f:D\to E$ in the category $\Cat{C}$ are mapped
  to an object and a  morphism in
  another category, $\Cat{Sets}$ in this case.
\item
  That arrow goes in the opposite direction, from the object associated with $E$
  to that associated with $D$.
\end{enumerate}
2. is easily addressed by constructing the category $\Cat{C^{op}}$ which has
the same objects as $\Cat{C}$, but whose arrows go in the opposite direction,
that is, instead of $f:D\to E$, we have $f^{op}:E\to D$. In some cases, this
looks very natural. For instance, a partially ordered set $(S,\le)$ suggests
the category whose objects are the elements $a,b,\dots$ of $S$, and the arrow
$a\to b$ means $a\le b$. The opposite category then corresponds to
$(S,\ge)$. Likewise, for $\Cat{Sets}$, the morphism $U\subset V$ becomes
$V\supset U$ in $\Cat{Sets^{op}}$. In other cases, the constuction of $\Cat{C^{op}}$ may seem
contrived, but, in fact, it will turn out to be quite important for the
structural theory.\\
To formalize 1., we state
\begin{defi}\label{def:Category4}A \textit{functor}
  \index{functor}$F:\mathbf{C} \to \mathbf{D}$ between two categories 
   maps objects and arrows of $\mathbf{C}$ to objects and arrows of
$\mathbf{D}$, respecting the category structures, i.e., for all objects $A,B$
and all arrows $f,g$ of $\mathbf{C}$, 
\begin{eqnarray}
\label{cat1-12}
&F(f:A\to B) \text{ is } F(f):F(A) \to F(B)\\
\label{cat1-13}
&F(g\circ f)=F(g)\circ F(f)\\
\label{cat1-14}
&F(1_A)=1_{F(A)}.
\end{eqnarray}
\end{defi}
Putting it differently and more abstractly, we consider the category
$\Cat{Cat}$ of all categories. Its objects are categories, and its morphisms
are functors. \\

With these concepts, the above construction can be formulated as
\begin{lemma}\label{yoneda1}
\begin{equation}
  \label{presh5}
  D\mapsto \Hom_\mathbf{C}(D,C)
\end{equation}  
defines a functor from $\Cat{C^{op}}$ to $\Cat{Sets}$.\end{lemma}

This functor depends on $C$, and we can therefore lift this to the next level
of abstraction.
\begin{equation}
  \label{presh6}
  C\mapsto \Hom_\mathbf{C}(-,C)
\end{equation}
maps each object $C$ to a functor. To formalize this, we introduce 
\begin{defi}\label{def:Category5}
The objects of the \textit{functor category}
$\mathbf{D}^{\mathbf{C}}$ are the functors $F:\mathbf{C}\to
\mathbf{D}$\index{category of functors}\index{functor category} between the categories $\mathbf{C}, \mathbf{D}$, The
morphisms of this category, called \textit{natural
  transformations}\index{natural transformations} 
\begin{equation}\label{cat1-x1}
\Theta:F\to G, 
\end{equation}
map a functor $F$ to a functor $G$, so that for the induced  morphism 
\begin{equation}\label{cat1-x2}
\Theta_C:FC\to GC
\end{equation}
 the diagram
\begin{equation}\label{cat1-x3}
\begin{CD}
    FC @>{\Theta_C}>> GC     \\
    @V{Ff}VV    @VV{Gf}V     \\
    FC' @>{\Theta_{C'}}>>GC'
  \end{CD}
\end{equation} 
commutes. That means in turn that
\begin{equation*}
  Gf\circ \Theta_C=\Theta_{C'}\circ FC'.
\end{equation*}
\end{defi}

\eqref{cat1-x3} is the same as \eqref{cat1-5b}, except that now we consider a
relation between morphisms in different categories.\\

With these notions in place, we can formulate the above construction as 
\begin{lemma}\label{yoneda2}
\begin{equation}
  \label{presh8}
  y:C\mapsto \Hom_\mathbf{C}(-,C)
\end{equation}  
defines a functor $y$ from $\Cat{C}$ to $\Cat{Sets^{C^{op}}}$.
\end{lemma}
\begin{defi}\label{def:Category6}
  The functor $y$ from  \eqref{presh8} is called the \textit{Yoneda functor}.
\end{defi}
We then have the \textit{Yoneda lemma}
\begin{lemma}\label{yoneda3}
   The morphisms between the functors
$\Hom_\mathbf{C}(-,C_1)$ and $\Hom_\mathbf{C}(-,C_2)$ correspond to the
morphisms between the objects $C_1$ and $C_2$ of the category $\Cat{C}$. 
In particular, the   objects $C_1,C_2$ are isomorphic if
 and only if  the functors $\Hom_\mathbf{C}(-,C_1)$ and
 $\Hom_\mathbf{C}(-,C_2)$ are isomorphic.
\end{lemma}
Let us write down the correspondance explicitly. The morphism $f:C_1\to C_2$
  induces $f:\Hom_\mathbf{C}(-,C_1)\to \Hom_\mathbf{C}(-,C_2)$ via
  \begin{equation}
    \label{presh10}
    (g:D\to C_1)\mapsto (f\circ g:D\to C_2). 
  \end{equation}
Thus, the simple relation \eqref{presh10} has enabled us to develop a general
construction and to formulate a general principle. \\
{\bf The objects of a category are characterized and determined by their
  relations with other objects.}

Again, the preceding constructions can be put into a more general framework.
\begin{defi}\label{Category6}A \textit{presheaf}\index{presheaf} on the category 
$\mathbf{C}$ is  an element $P$ of
the functor category $\mathbf{Sets^{C^{op}}}$.
\end{defi}
Thus, a presheaf is a functor $P:\mathbf{{C^{op}}}\to \mathbf{Sets}$.\\
When going from  an arrow $f:V \to U$ in $\mathbf{C}$ to its image  $Pf:PU\to
PV$, we reverse the  direction, going  from
$\mathbf{C}$ to $\mathbf{{C^{op}}}$.
\begin{defi}\label{Category7}Let $P$ be a presheaf on $\Cat{C}$. Let $f:V \to U$ be an arrow
  in $\Cat{C}$. For  $x\in PU$ , we call the value
$Pf(x)$ 
the \textit{restriction}\index{restriction} of $x$ along $f$. 
\end{defi}

Thus, the Yoneda functor is a presheaf. And we can now formulate the following
more general version of the \textit{Yoneda lemma}.
\begin{lemma}\label{Yoneda4}Let $F\in \mathbf{Sets^{C^{op}}}$ be a presheaf on the category
  $\Cat{C}$. Then for any object $D$ of $\Cat{C}$, we have the isomorphism
  \begin{equation}
    \label{presh11}
    \Hom_{\Cat{Sets^{C^{op}}}}(yD,F)\cong FD.
  \end{equation}
  \end{lemma}
 This isomorphism behaves naturally both w.r.t. the object $D$ and the functor
 $F$, that is, changes of $D$ or $F$ lead to natural commutative diagrams, as
 both of them appear on both sides of \eqref{presh11}.\\
 In particular, if we take $F=yC$, then $FD=yC(D)=\Hom_{\Cat{C}}(D,C)$, the
 statement of Lemma \ref{yoneda3}.\\

 Let us also record
   \begin{lemma}\label{prefun}
     Let $F:\Cat{C}\to \Cat{D}$ be functor. Then it induces a functor
     \begin{equation}
       \label{presh12}
       F^\ast:\Cat{Sets^{D^{op}}} \to \Cat{Sets^{C^{op}}}.
     \end{equation}
   \end{lemma}
   For the \textit{proof}, we simply put for a presheaf $Q\in \Cat{Sets^{D^{op}}}$ on $\Cat{D}$ and $C\in \Cat{C}$
   \begin{equation}
     \label{presh12a}
     F^\ast Q(C)=Q(FC)\ .
   \end{equation}
\bigskip

The concept of a presheaf is useful much beyond
that. A presheaf can be considered as a family of sets, indexed by the objects
of the category $\Cat{C}$, in such a way that we can restrict via a morphism
$f:U\to V$ from $V$ to $U$. In particular, when we have two such index sets
$A(C)$ and $B(C)$, we can ask for which $C$ we have
\begin{equation}
  \label{presh13}
  A(C)\subset B(C).
\end{equation}
Again, this can be abstracted and formalized, leading to the notion of a
topos.\\

Here, we want to consider a more concrete situation. Let $U$ be a set, perhaps
with some additional structure, like that of a topological space. We construct
a presheaf $P$ by assigning to each (open) subset $V\subset U$ the set of
(continuous) functions on $V$. The restriction property is satisfied, because
the restriction of a (continuous) function to an (open) subset is again a
(continuous) function. A \textit{section} of this presheaf then  then is an
element of
$PU$, that is, a (continouus) function on
$U$. The restriction property of the presheaf then implies that for each
$V\subset U$, the restriction of the section to $V$ then likewise is an
element of $PV$, that is, a (continuous) function on $V$. Conversely, for $V\subset U$, we can likewise construct such local sections. In the
topological case, such a local section need not extend to a global one,
because a continuous function on $V$ may not be extendable to a continuous
function on $U$.\\

\bigskip

So, how can we use this quite abstract framework for our purposes?
The basic category $\Cat{C}$ stems from  our set $S$ of features or properties, possibly
equipped with some additional structure beyond just being a set. There could be a specific type of relations
between the elements $p,q,r,\dots $ of such a set.  A set with such a structure could for instance be a graph, with a morphism $p\to
q$ if there is a directed edge from $p$ to $q$. It could be symmetric, that
is,  $p\to
q$ precisely if $q\to p$, as in some infrastructural networks. It could also
be a metric space, that is, there could be a distance function $d(.,.)\ge 0$, with
$d(p,q)=0$ only if $p=q$, $d(p,q)=d(q,a)$ and $d(p,q)\le d(p,r)+d(r,q)$ for
all $p,q,r$. More generally, there could be a similarity relation
$\sigma(.,.)$ , with $0\le \sigma(p,q)\le 1$, which is again symmetric,
$\sigma(q,p)=\sigma(p,q)$ and $\sigma(p,q)= 1$ only for $p=q$ and
$\sigma(p,q)=0$ when $p$ and $q$ are completely dissimilar. The category
$\Cat{C}$, however, would not be such a set itself, but some class of subsets
of it, with morphisms $A\to B$ standing for the subset relation $\subset
B$. If $S$ is simply a set, then any subset of $S$ would be an object of our
category, but when it carries a metric or a similarity structure, we would
start with the open balls $U_r(p)=\{q: d(p,q) <r\}$ or $U_r(p)=\{q:
\sigma(p,q) > 1-r\}$ for some $r$, and their finite intersections and
arbitrary unions (to create what is called a topological space). But in
principle, we can take as our category $\Cat{C}$ any collection of subsets of
$S$, ordered by inclusion and selected according to properties that we like to
consider.\\
Since category theoretical constructions can be iterated and lifted to higher
levels, we can also consider the categories of all sets, or of all metric or
similarity structures. Then a morphism between sets is an arbitrary map
between sets. A morphism between two metric structures $(S_1,d_1)$ and
$(S_2,d_2)$ would be a map $f:S_1\to S_2$ with $d_2(f(p),f(q)\le d_1(p,q)$ for
all $p,q\in S_1$. Likewise a morphism between similarity structures would need
to satisfy $\sigma_2(f(p),f(q))\ge \sigma_1(p,q)$ for all $p,q$.\\
Later on, we shall even consider categories of presheaves.\\

But let us first explain why presheaves are useful and efficient models of
systems that can the be modified and combined for innovations. Basically,
we have a set $S$ of features, possibly with some structure. And the category
$\Cat{C}$ that we shall first work with is one of subsets of $S$, as just
explained. A presheaf $\Cat{P}$ then assigns to each such $V\subset S$ a
set. We interpret that set as the collection of compatible feature
values. When $V$ is a singleton, that is, stands for a single feature, then
$P(V)$ is the set of all possible values that that feature can assume. When
$V$ consists of two elements, $p,q$, then $P(V)$ is the set of all feature
values of $p$ and $q$ that are compatible with each other. For instance, when
in our organization model, the feature $p$ stands for size and $q$ for
hierarchical levels, then a small size and a large number of hierarchical
levels are not compatible with each other and therefore not contained in the
set $P(V)$. In contrast, large size is compatible with many hierarchical
levels, but perhaps also with few of them, as is the case for flat
hierarchies.
Thus, in the simplest case, we work with two sizes $l,s$ for large and small,
and two level values, $m,f$ for many and few, and $P(V)$ contains the pairs
$(s,f), (l,m), (l,f)$, but not $(s,m)$. Of course, in practice, we shall allow
for more feature values, perhaps positive integers, but the principle should
be clear from this impoverished example. \\
The important point is that the larger the feature set, the more compatibility
conditions have to be satisfied between the values for the individual
features. Thus, when $U\subset V$, we can restrict the feature combinations in
$PV$, that is, those 
that are valid for $V$ to $U$, and we then get feature combinations that
satisfy all the compatibility constraints on $U$, that is, are in
$PU$. Therefore, we have indeed a presheaf. \\

Now, when $V$ is large, there are typically only relatively few feature
combinations that satisfy all the compatibility conditions between all the
feature values in $V$. Therefore, $PV$ then is small, and this is good,
because then we only need to take relatively few possible combinations into
account and can discard all others. \\

\medskip 

To illustrate these formal structures, let us consider an {\bf example}. For the presentation of the example, we shall also use the concept of a \textit{section} which will be formally introduced in Section \ref{sec}.\\
We describe \textit{countries} by their features, geographic ones like size, average elevation, coast length, ..., demographic ones, like population size, birth and death rates, life expectancy, age or gender distribution, health status, main causes of illness, ..., economic ones, like gross national product, average or mean income, income distribution, tax rates, unemployment rate, employees in different economic sectors,..., social ones like  education, social inequality, city, town and countriside population, level of happiness, ... or whatever. Not only the individual features, but also certain sets of features are meaningful and used in statistics. But not all of them would be used. For instance, it might not make much sense to only record the coast length, the number of people in the agricultural sector and the average level of happiness. But one might take the set of all demographic features only. Thus, there are selected subsets $V$ of the total  feature set $S$. These $V$ are ordered by inclusion. For instance, if $V_0$ is the set of all economic indicators, we can take the subset $V_1$ containing only the gross national product , the income distribution and the unemployment rate. Thus, $V_1\subset V_0$ represents a morphism in our category of specific subsets of $S$. For the opposite category, $V_0 \supset V_1$ would represent an arrow. \\
Some features take simple scalar values. For instance, the gross national product can be expressed in Euros, and the average life expectancy in years and months. Others, like the age distribution take more complex values, while the level of happiness may perhaps assume only some few discrete ones, like high, medium and low. When assigning to each $V$ in our category the set $P(V)$ of values that their individual features can possibly take, we get a presheaf $P$. And a country $C$ then specifies a section, by assigning to each feature the value that it possesses for $C$. When we ignore some of the features, or if for some country, some values are not available, we obtain a local section. And a global section can always be restricted to a local one, by ignoring the values of some features that may not be of interest for the purpose at hand. If we take the values of sufficiently  many features, the corresponding combination of values, that is, the section of the presheaf, specifies a country uniquely. But if we take only few ones, for instance only the level of happiness, then many countries will have the same value, and so, that value no longer determines the country. Also, many value combinations are not assumed by any country. For instance, a high gross national product is not compatible with a small population with a low average income. Putting this more positively, when the gross national product is high and the population small, then we can conclude that average income is high. In other words, there are correlations between the feature values that can be assumed for any country, and this allows us to make certain inferences without having to measure the values of all features in question.

\section{Operations on presheaves and the reorganization and combination of
  structures}

\subsection{Diagrams} 
We consider the functor category $\Cat{A^I}$, that is, the category of
functors from $\Cat{I}$ to $\Cat{A}$.  where $\Cat{A}$ in our applications
below, usually is a category of subsets of some set $S$, with morphisms being
inclusions, and where $\Cat{I}$ is a  category  with finitely many objects and morphisms, and it is then called an
\textit{index category}. For the category  $\Cat{2}$ with two objects $1,2$ and
the morphism $1\to 2$, a functor $\Cat{2}\to \Cat{A}$ selects a morphism
$A_1\to A_2$ in $\Cat{A}$. In general, a functor maps an object $i$ in
$\Cat{I}$ to the object $A_i$ in $\Cat{A}$ in $\Cat{A}$ and a morphism $i\to j$ in
$\Cat{I}$ to the morphism $A_i\to A_j$ in $\Cat{A}$. We  call the image of
such a 
functor, that is the pattern consisting of the objects $A_i$ and the morphisms
$A_i\to A_j$ in $\Cat{A}$ a \textit{diagram}.

When we consider the index category $ \mathbf{I}$ with the
  structure given by the commutative diagram 
  \begin{equation}\label{cat0}
\begindc{\commdiag}[20]
\obj(50,30)[2]{$2$}
\obj(30,30)[1]{$1$}
  \obj(30,10)[3]{$3$}
  \obj(50,10)[4]{$4$}
  \mor{1}{2}{}
  \mor{1}{3}{}
   \mor{3}{4}{}
  \mor{2}{4}{}
  \enddc
\end{equation} 
 a  diagram in $\Cat{A}$ 
 is
 \begin{equation}\label{cat0a}
\begindc{\commdiag}[20]
\obj(50,30){$A_2$}
\obj(30,30){$A_1$}
  \obj(30,10){$A_3$}
  \obj(50,10){$A_4$}
  \mor{$A_1$}{$A_2$}{}
  \mor{$A_1$}{$A_3$}{}
   \mor{$A_3$}{$A_4$}{}
  \mor{$A_2$}{$A_4$}{}
  \enddc
  \end{equation}

\begin{defi}\label{Diagram1}
  A \textit{cone} over a diagram in $\Cat{A}$ is an object $C$ of $\Cat{A}$ with arrows 
\begin{equation}\label{cat1}
a_i:C \to A_i,
\end{equation}
leading  for each arrow  $i\to j$ in $\mathbf{I}$ to the commuting diagram 
\begin{equation}\label{cat2}
\begindc{\commdiag}[20]
  \obj(30,30){$C$}
  \obj(10,10){$A_i$}
  \obj(50,10){$A_j$}
  \mor{$C$}{$A_i$}{$a_i$}[\atright,\solidarrow]
  \mor{$C$}{$A_j$}{$a_j$}
  \mor{$A_i$}{$A_j$}{}
  \enddc
  \end{equation}
  And a \textit{cocone} is an object $D$ with arrows
  \begin{equation}\label{cat1a}
b_i: A_i\to D
\end{equation}
and commuting diagrams
\begin{equation}\label{cat2a}
\begindc{\commdiag}[20]
  \obj(30,10){$D$}
  \obj(10,30){$A_i$}
  \obj(50,30){$A_j$}
  \mor{$A_i$}{$D$}{$b_i$}[\atright,\solidarrow]
  \mor{$A_j$}{$D$}{$b_j$}
  \mor{$A_i$}{$A_j$}{}
  \enddc
  \end{equation}
\end{defi}
\begin{defi}\label{Diagram2}
  A \textit{limit}\ $p_i:C_{\to}\to A_i,
i\in \mathbf{I}$, for a diagram  is a cone that satisfies the universal 
 property that for each cone  $(C,a_i)$ over that diagram there is a unique
morphism
 $c:C\to C_{\to}
$ with
\begin{equation}\label{cat7}
p_i\circ c =a_i \text{ for all }i.
\end{equation}
Analogously, we define a \textit{colimit} in terms of such a universal property for cocones. 
\end{defi}

When $\Cat{I}$ has two objects $1,2$ and no non-trivial morphisms, a cone over
a diagram is an object $C$ with  arrows
\begin{equation}\label{cat22}
D_1\begin{CD}@< c_1 << \end{CD}C \begin{CD}@>c_2>> \end{CD} D_2,
\end{equation}
A limit of such a diagram is called a \textit{product} of $D_1,D_2$.\\
For the  index category  $\mathbf{I}=\{1,2,3\}$
\begin{equation}\label{cat22f}
\begindc{\commdiag}[20]
  \obj(50,30){$1$}
  \obj(30,10){$2$}
  \obj(50,10){$3$}
  \mor{$1$}{$3$}{}
  \mor{$2$}{$3$}{}
  \enddc
\end{equation} 
a limit of a corresponding diagram
\begin{equation}\label{cat22faaa}
\begindc{\commdiag}[20]
  \obj(50,30){$D_1$}
  \obj(30,10){$D_2$}
  \obj(50,10){$D_3$}
  \mor{$D_1$}{$D_3$}{$d_1$}
  \mor{$D_2$}{$D_3$}{$d_2$}
  \enddc
\end{equation} 
is  called a {\it pullback}\index{pullback} of $d_1, d_2$. Again, dual
constructions are possible and called \textit{coproducts} and
\textit{pushouts}. Neither of these need to exist in a given category, but in
those that we shall consider  they do. \\
In fact, the category of interest for us is the power set $\mathcal{P}(S)$ of
some (finite) set $S$, that is, the set of all subsets of $S$, or some
subcategory of this category, In that category, the product of $S_1,S_2\subset
S$ is the intersection $S_1\cap S_2$, and the coproduct is the union $S_1\cup
S_2$. Thus, when we have sets $S_1,S_2\subset S$, we get the diagram 
\bel{cat220fae}
\begindc{\commdiag}
  \obj(30,30){$S_1\cap S_2$}
  \obj(50,30){$S_1$}
  \obj(30,10){$S_2$} 
  \obj(50,10){$S_1\cup S_2$}
  \mor{$S_1$}{$S_1\cup S_2$}{}
  \mor{$S_2$}{$S_1\cup S_2$}{}
  \mor{$S_1\cap S_2$}{$S_1$}{}
  \mor{$S_1\cap S_2$}{$S_2$}{}
  \enddc
\qe
in which $S_1\cap S_2$ is a pullback and $S_1\cup S_2$ is a pushforward.

\subsection{Categories of power sets}\label{powerset}
We consider a finite, but typically rather large set $S$, standing for all
possible features. We let $\mathcal{P}(s)$ be the power set of $S$, that is,
the set of all the subsets of $S$. When we let $U\to V$ stand for $U\subset
V$, it becomes a category. We also let $\mathcal{U}(S)$ be some subcategory of
$\mathcal{P}(s)$  that is closed under intersection and union, that is, if
$U,V\in \mathcal{U}(S)$, then so are $U\cap V$ and $U\cup V$. For instance,
$S$ could be a topological space, and $\mathcal{U}(S)$ be the set of its open
subsets. For a subset $S_0\subset S$, we let
$\mathcal{U}(S_0)=\mathcal{P}(S_0)\cap \mathcal{U}(S)$, the set of all subsets
of $S_0$ that belong to $\mathcal{U}(S)$. Each such $\mathcal{U}(S_0)$ then is
a set that is partially ordered by inclusion, and again, we consider it as a
category with the morphisms being inclusions.

We then let $\Cat{C}$ be the
category whose objects are the categories $\mathcal{U}(S_0)$ for $S_0\subset
S$. The morphisms are functors $f:\mathcal{U}(S_1)\to \mathcal{U}(S_2)$. As a function, $f$ of course has to
preserve inclusions, that is, if $U\subset V$, then $f(U)\subset f(V)$. If we
alos want to have $U\subset f(U)$ for all $U\subset S_1$, then  $S_1\subset
S_2$, and there are two extremal
possibilities for such a functor $f$:
\begin{equation}
  \label{cat55}
  f_0(U)=U
\end{equation}
and
\begin{equation}
  \label{cat56}
  f_1(U)=U\cup  (S_2\backslash S_1)\ .
 \end{equation} 
  We also have the restriction functor
  \begin{eqnarray}
    \nonumber 
    r:&\mathcal{U}(S_2)&\to \mathcal{U}(S_1)\\
    \label{cat50}
    &U&\mapsto U\cap S_1
  \end{eqnarray}
  for $S_2\subset S_1$. We can also consider $r$ as a morphism between the
  opposite categories $r:\mathcal{U}(S_2)^{op}\to \mathcal{U}(S_1)^{op}$ with
  $S_2\supset S_1$. 
  \begin{lemma} Let $S_1\subset S_2$. 
  $r$ is the right adjoint of $f_0$ and the left adjoint of $f_1$.   
  \end{lemma}
  \begin{pf}
    $L:\Cat{C} \leftrightarrow \Cat{D}:R$ are adjoint to each other if
    \begin{equation}
      \label{cat60}
      \Hom_{\Cat{D}}(LC,D)=\Hom_{\Cat{C}}(C,RD)
    \end{equation}
    for all objects $C$ of $\Cat{C}$ and $D$  of $\Cat{D}$. We have
    \begin{equation*}
      \Hom_{\mathcal{U}(S_2)}(U,V)= \Hom_{\mathcal{U}(S_1)}(U,V\cap S_1)
    \end{equation*}
     and
   \begin{equation*}
      \Hom_{\mathcal{U}(S_1)}(V\cap S_1,U)= \Hom_{\mathcal{U}(S_2)}(V,U\cup
      (S_2\backslash S_1))
    \end{equation*}
 for $U\subset S_1, V\subset S_2$ because $U\subset V$ iff $U\subset V\cap
 S_1$ and    $V\cap S_1 \subset U$ iff $V \subset U\cup
      (S_2\backslash S_1)$. 
  \end{pf}\qed\\
  We then look at commutative diagrams of the form
  \bel{cat65}
\begindc{\commdiag}[50]
\obj(30,10)[0]{$\mathcal{U}(S_1\cup S_2)$}
\obj(10,20)[1]{$\mathcal{U}(S_1)$}
\obj(30,30)[2]{$\mathcal{U}(S_1\cap S_2)$}
\obj(50,20)[3]{$\mathcal{U}(S_2)$}
\mor{1}{0}{}
 \mor{3}{0}{}
 \mor{2}{1}{}
 \mor{2}{3}{} ,
\enddc
\qe
When the morphisms are either of the form $f_0$ \rf{cat55} or of the form $f_1$ 
\rf {cat56}, then $\mathcal{U}(S_1\cap S_2)$ is a pullback and
$\mathcal{U}(S_1\cup S_2)$ is a pushout.\\
We also have the commutative diagram
 \bel{cat66}
\begindc{\commdiag}[50]
\obj(30,10)[0]{$\mathcal{U}(S_1\cap S_2)^{op}$}
\obj(10,20)[1]{$\mathcal{U}(S_1)^{op}$}
\obj(30,30)[2]{$\mathcal{U}(S_1\cup S_2)^{op}$}
\obj(50,20)[3]{$\mathcal{U}(S_2)^{op}$}
\mor{1}{0}{}
 \mor{3}{0}{}
 \mor{2}{1}{}
 \mor{2}{3}{} 
\enddc
\qe
where we now let the morphisms be the restriction maps \rf{cat50}.\\

Now let $Q\in \Cat{Sets}^{\mathcal{U}(S)^{op}}$ be a presheaf on
$\mathcal{U}(S)$. We then  have the commutative pullback-pushout diagram
 \bel{cat67}
\begindc{\commdiag}[50]
\obj(30,10)[0]{$Q(S_1)\cup Q(S_2)$}
\obj(10,20)[1]{$Q(S_1)$}
\obj(30,30)[2]{$Q(S_1)\cap Q(S_2)$}
\obj(50,20)[3]{$Q(S_2)$}
\mor{1}{0}{}
 \mor{3}{0}{}
 \mor{2}{1}{}
 \mor{2}{3}{} 
\enddc
\qe
By the presheaf property, $Q(S_1\cup S_2)$ restricts to $Q(S_1)\cap Q(S_2)$,
and $Q(S_1\cap S_2)$ contains the restrictions of  $Q(S_1)\cup Q(S_2)$.\\

The preceding now suggests a merging operation on presheaves.
\begin{defi}\label{premerge}
  Let $Q_1$ be a presheaf on $\mathcal{U}(S_1)$, and  $Q_2$ be one on
  $\mathcal{U}(S_2)$. We then \textit{merge} them as a presheaf $Q$ on
  $\mathcal{U}(S_1 \cup S_2)$ by letting $Q(S_1\cap S_2)=Q_1(S_1\cap S_2)\cup
  Q_2(S_1\cap S_2)$ (more precisely, using this construction on every $S_0\subset
  S_1\cap S_2$), and be using $Q_1$ on $S_1\backslash  S_1\cap S_2$ and  $Q_2$
  on $S_2\backslash  S_1\cap S_2$. 
\end{defi}

The question then is which of the elements of $Q(S_1\cap S_2)$ can be extended
to both $S_1$ and $S_2$, and hence to $S_1\cup S_2$.  And this is what we shall study in our applications
below.\\

Of course, there are also some trivial operations on presheaves. For instance,
we could try to enlarge $Q(S_0)$ for some $S_0\subset S$. That is, we extend
the range of $Q$ on $S_0$. In applications, this then might also make it
possible to also extend $Q(S)$. \\
Or we could shrink or enlargen $S$ itself. In the first case, there are then
fewer constraints for global extensions, and in the second case, more. All
this is fairly obvious and will not be treated in detail.

\section{Applications to business innovations. The Apple case}

\subsection{Sections}\label{sec} 
We consider again a set $S$ of features, and a subset $\mathcal{U}(S)$ of its
power set $\mathcal{P}(S)$. As before, $\mathcal{U}(S)$ should be closed under
intersections and unions, and we also assume that it contains all single
element sets $\{x\}$ for $x\in S$. And as before, for $S_0\subset S$,
$\mathcal{U}(S_0)=\mathcal{U}(S)\cap \mathcal{P}(S_0)$. Let $Q\in
\Cat{Sets}^{\mathcal{U}(S)^{op}}$ be a presheaf on $\mathcal{U}(S)$.
\begin{defi}
  A \textit{global section} $s$ of the presheaf $Q\in \mathcal{U}(S)$ is an
  $s\in Q(S)$. By the restriction property, it then  assigns to
  every $x\in S$ an element  $s(x)\in Q(x)$ (where $Q(x)$ is an abbreviation of
  $Q(\{x\})$). 
   A \textit{local section} $s_0\in Q(S_0)$ on $S_0\subset S$ similarly assigns to
  $x\in S_)$ some $s_0(x)\in Q(x)$. 
\end{defi}
\begin{lemma}
  Restricting a glocal section $s$ to $S_0\subset S$ yields a local section on
  $S_0$. 
\end{lemma}
\begin{pf}
  By the presheaf property, restricting an element of $Q(S)$ to $S_0$ yields an
  element of $Q(S_0)$.
\end{pf}\qed\\
 The converse, however, does not necessarily hold. A local section need not
 extend to a global one. \\
 
This now is the important point for our applications. A (local) section $s$ selects a
value for each feature $x$ in our base set $S_0$. The feature values need to
be compatible with each other, as dictated by the requirement that $s\in
Q(S_0)$/ In other words, the presheaf $Q$ encodes the compatibility
requirements between feature values. The larger $S_0$, the more requirements
there are, and correspondingly, the fewer sections we have. Restricting from
$S_0$ to some $S_1\subset S_0$ also yields a restricted section $s_{|S_1}$,
but in contrast, a local section on $S_1$ need not extend to one on a larger 
set $S_0$. \\

This may be due, for example, to limitations in the range of values of the new features considered when trying to extend a section. For example, Henry Ford (in cooperation with Edison) seriously tried to develop in 1914-16  an affordable electric vehicle, but the project was abandoned because, given that the range of values of size, weight, power and duration of batteries in those years turned out to be incompatible with the features of an affordable vehicle \cite{Bryan}.

 \begin{figure}[h]\label{canvas}
   \centering
   \includegraphics{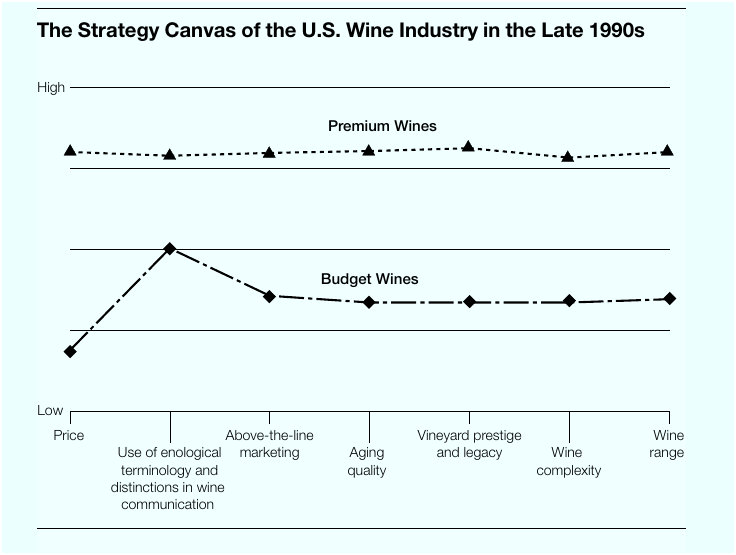}
   \caption{Strategy canvas of the U.S. wine industry in the late 1990s, from
   \cite{KM}}
 \end{figure}

Presheaves and their sections
provide an efficient way of representing constraints for combinations of feature values. Related graphic representation are sometimes used in the strategic literature - an example is the strategy canvas of \cite{KM} in Fig.\ref{canvas}  In this example, we have seven features, and representing the possible values in a Cartesian space $\R^n$ would need $n=7$, much too large
for a Cartesian graphical representation. \\
In the Fig. \ref{canvas}, we see two (global) sections, one corresponding to premium wines, where the values of all features are high, and the other for budget wines, with consistently lower values. Importantly, there is no section for which some values are high while others are low, as this would not yield a viable business strategy. Or, putting it the other way around, if we already know that for a certain strategy, some values are low, we can expect that the others are low as well.

\subsection{The digital hub}
\begin{figure}[h]
  \centering
  \includegraphics{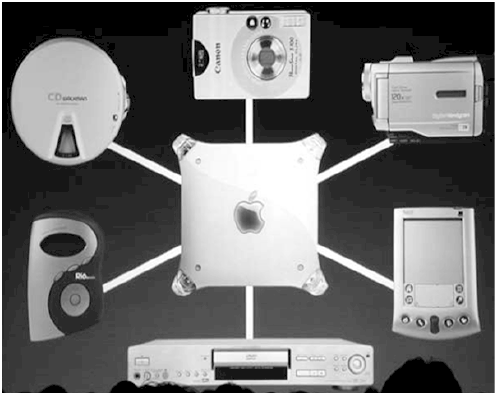}
  \caption{The digital hub}
  \label{hub}
\end{figure}

The current section  contains the case study on the digital hub (originally contributed by MW) that was also presented in a former unpublished working paper - here it is more tightly integrated into the mathematical formalism. \\
In one of his most well-known keynote presentations \cite{Jo}, in Jan. 9  2001, Steve Jobs
delineated a strategic concept, the digital hub, that proved fundamental for
the spectacular ascent of Apple to prominence in the market for digital
devices \cite{Jo}. 
In contrast to a dominant trend towards considering the PC waning, Jobs drew  a vision of an explosion of a world of devices rotating around the PC, making it the "digital hub" of the emerging digital lifestyle. Due to its computational power, large and inexpensive storage, big screen, connectivity to the internet, the PC could add value to digital devices and interconnect them. A graphical depiction of the digital hub, taken from Steve Job's own presentation, is reproduced in figure \ref{hub}. In what follows, we will analyze the conceptual structure underlying the digital hub using the concepts and formal tools we introduced above.
- -

\subsubsection{Amalgamation}

The seed of the digital hub concept, in the words of Steve Jobs, can be found
in the creation of the iMovie 2 application, that multiplied the functional
possibilities associated to a camcorder. We will now discuss the conceptual
moves underlying the 'iMovie' strategy using our formal language. We shall use
merge procedures explained in Section \ref{powerset}. In our present context,
we shall speak of an \textit{amalgamation}. And instead of features, we shall
sometimes speak of \textit{observables}. The two words mean the same thing
here. And the set of possible values of such an obversable we will also call a \textit{fiber}. In our example, the values will be discrete, and therefore, each such fiber is a finite set. \\

We consider  two presheaves $Q_1,Q_2$ with base sets $O_1,O_2$. $O_1$ contains the subobservables of the observable 'PC' that matter for connecting it with video devices. $O_2$ contains the single observable camcorder ('$V$') and its subobservables. Figure \ref{imovie} describes the observables in both 'PC' and 'Camcorder'. These are the black labels on the fibers over PC and Camcorder. Concretely, PC contains the observables 'film content', 'screen size', and 'computing power'. Camcorder contains the observables 'film content', 'screen size', and 'possibilities to edit content captured by the camera'. The observables 'screen size' and 'film content' appear in both PC and Camcorder and can be amalgamated when we consider the possibilities of PC and Camcorder together. This is an example of the merge operation abstractly given in Def. \ref{premerge}.

\begin{figure}[h]\label{imovie}
  \centering
  \includegraphics[scale=.5]{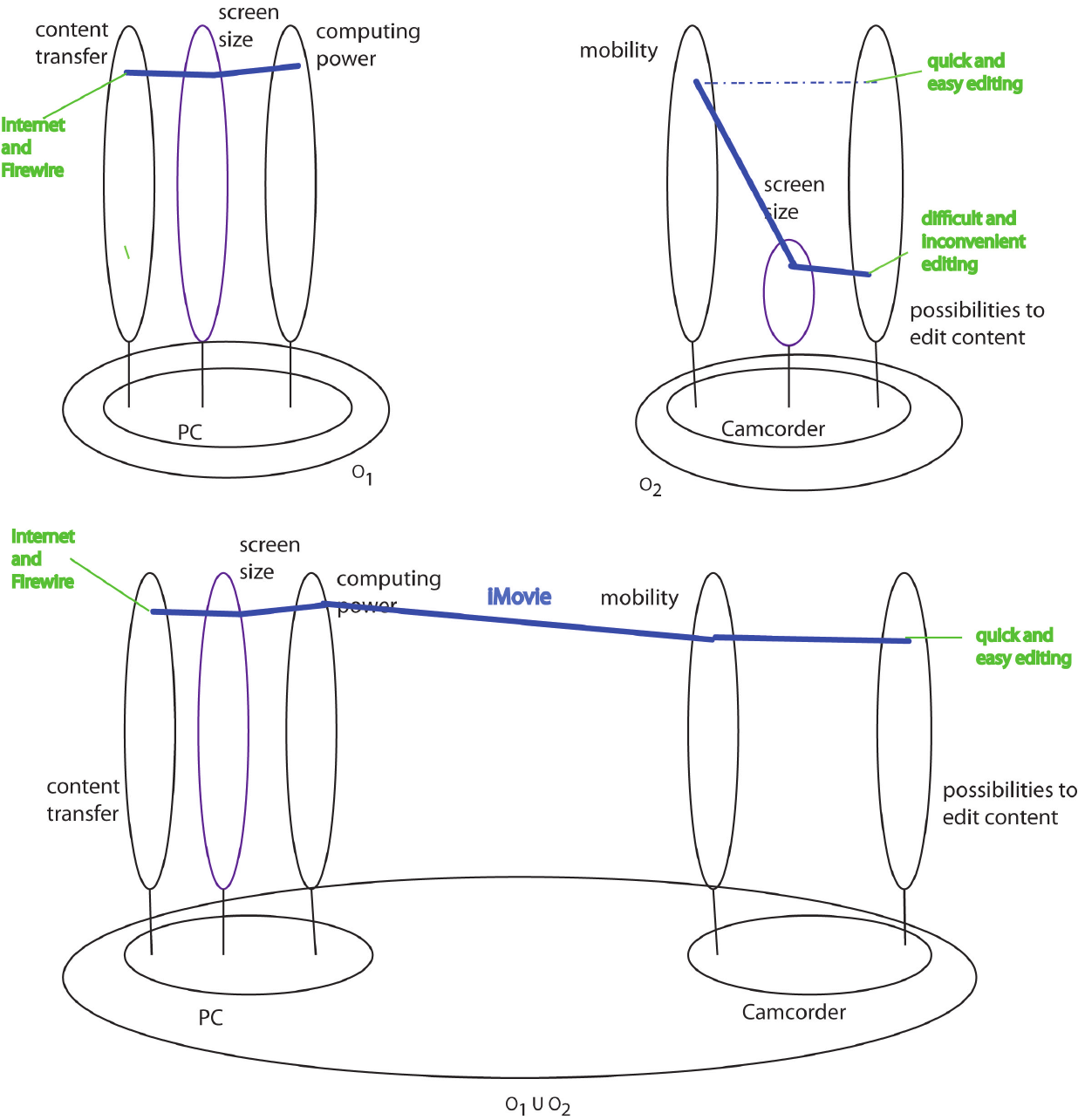}
  \caption{iMovie: a pictorial representation of the amalgamation of PC and Camcorder}
  \label{imovie}
\end{figure}

A PC of course permits a larger screen size than a videocam. We then
define the amalgamation of the two presheaves as the presheaf with
base set $O_1 \cup O_2$, and with the value ranges of each observable
$x \in O_1 \cap O_2$, that is, $x$ occurs in both $O_1$ and $O_2$ as
the screen size in the above example, being the union $Q_1(x) \cup Q_2(x)$
of the individual value ranges. Thus, the amalgamation of the PC and
the videocam makes a larger screen size available for processing video
data. And since, when as in our example $Q_2 \text{  } screen size$ is smaller than $Q_1 \text{  } screen size$,
by this amalgamation the value range for $x \in O_2$ gets extended,
then, as explained above, new global sections could
emerge. \\

Figure \ref{imovie} illustrates the principle
of merging or amalgamation by combining elements, and how this affects the
constraints of those elements. Consider the local section in $O_2$ in figure \ref{imovie}. The observable 'screen size' constraints the admissible values of the observable 'possibilities to
edit content captured by the camera'. The section that runs through all three observables of $O_2$ does not run through 'quick and easy editing'. However, there is a local section that connects the observables 'film content' and 'possibilities to edit content
captured by the camera' which runs through 'quick and easy editing'. Local sections that don't extend to global sections are expressed by dotted blue lines. The possibility of quick and easy editing is given in principle, it is just ruled out by the constraints imposed by the observable 'screen size'.
By means of amalgamation of 'screen size' in PC and Camcorder this constraint can be overcome. In $O_1 \cup O_2$ the global section can now run through 'quick and easy editing'.

In blue, we labeled some concrete sections, coherent
configurations of values for the observables.  In green, we
labeled some concrete assignments of values in
the sections. In $O_1$, the value in 'film content' is 'professional
content only', because before the days of iMovie, it was expensive to
edit films on digital platforms. Back then, film editing on computers
was dominantly done by professionals.

The amalgamation of PC and videocam also has the consequence
  that the effective range of film content on the side of the PC
  is enlarged.  The amalgamation of PC and Video makes it more likely that amateur content is edited on a PC.

We now describe this example in formal terms. The presheaf $QO_1$ is described as follows (it
  contains just one section here, but of course could contain more):
\bel{FO_1}
Q O_1 = \left\{
\begin{array}{c}
\text{professional film content only }\\
\text{screen size is large}\\
\text{computing power is large}
\end{array}
 \right\}
\qe
$Q O_2$ is described as follows:
\bel{FO_2}
Q O_2 = \left\{
\begin{array}{c}
\text{professional and amateur film content}\\
\text{screen size small}\\
\text{difficult and inconvenient editing}
\end{array}
\right\}
\qe

These formal definitions correspond to the blue sections in figure \ref{imovie}.

We here also describe the presheaf over the two observables 'film content' (which we denote $o_2$) and 'possibilities to
edit content captured by the camera' (which we denote $o_3$). An available local section expresses that easy possibilities to edit content are available in principle; they are just constrained by the screen size:
\bel{FsubO2}
Q \{ o_2, o_3 \} = 
\nonumber \\
\left\{
\begin{array}{c}
\text{professional and amateur film content}\\
\text{quick and easy editing}
\end{array}
\right\}
\qe

After amalgamation, we get the presheaf with base $O_1 \cup
  O_2$. Note that here those observables that are common to both $O_1$
  and $O_2$ have been identified. In particular, there now is only a
  single observable screen size, with possible values {\it large} and
  {\it small}, whereas in $O_2$, only the value {\it small} had been
  permitted. Because of this fact, we now have 
a new section (we omit the combination of the original sections in
\ref{FO_1} and \ref{FO_2} as they are no longer of interest).
\begin{eqnarray}
  \label{imoviesection}\nonumber
 & Q(O_1\cup O_2) \supset&  \qquad \qquad \qquad \qquad \qquad \\
  \nonumber
  \{ &\left(\begin{array}{c}
\text{amateur and professional editing on a computer}\\
\text{screen size is large}\\
\text{computing power is large}
\end{array}\right)& ,\\
&\left(\begin{array}{c}
\text{amateur and professional content }\\
\text{screen size small}\\
\text{quick and easy editing}
\end{array}\right)&   \}
\end{eqnarray}

Note that once value ranges are extended, it is not necessarily true that \textit{any} new global section would be possible.  The presheaf condition, which imposes a constraint preservation criterion, says that the section in
  $Q(O_1 \cup O_2)$ needs to run through points of existing local sections in all dimensions whose value ranges have not been
  extended. As already discussed, in figure \ref{imovie}, the dotted line in $O_2$ indicates a local section. The section in $Q(O_1\cup O_2)$ runs through 'amateur and professional content' and 'quick and easy editing' just as the local section that we defined above runs through these values.

Thus, the principles of constraint preservation and amalgamation are combined. Indeed, in any operation of amalgamation that leaves some dimensions as they were, the constraint preservation criterion also imposes constraints on possible emerging coherent value combinations.\\

\subsubsection{Transferring structure}

\begin{figure}[h]
\centering
\caption{\textbf{The structure underlying Apple's "Digital hub" wheel} }
\includegraphics[scale=0.5]{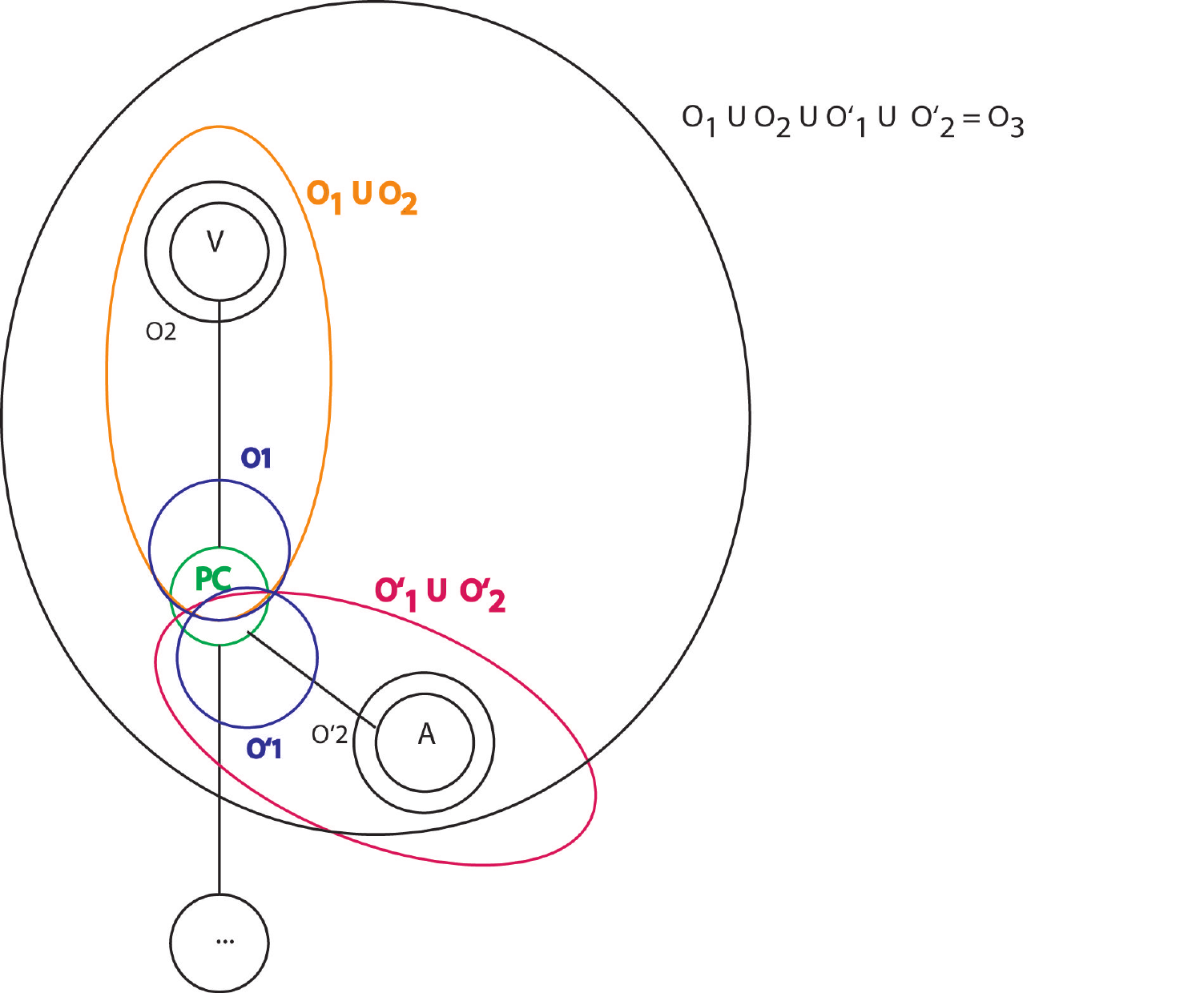}
\label{strucofhub}
\end{figure}

The digital hub is constructed via a series of analogies that expand the wheel
and of integrations that glue together its pieces. (Fig. \ref{hub} illustrates
that the structure of the digital hub resembles a wheel.) For instance, 2001
was the year of the introduction of iTunes and the iPod. We will demonstrate
that the iTunes strategy was structurally similar to the iMovie strategy and
express this similarity formally. To analyze the case formally, we first
introduce some notation, which is summarized in Fig. \ref{strucofhub}.

 We  denote  sets of observables as $O_1, O_2, \dots$. 'Sets of observables' contain observables of a strategy considered together. They can contain just one observable (like 'camcorder') and their subobservables or a collection ('camcorder' and 'PC') and their subobservables. 

$Q(O_1 \cup O_2)$ contains the constraints of considering possibilities of PC and Camcorder (Video) together: $O_1 \cup O_2 =\{PC, V\}$.  $F O'_1 \cup O'_2$ contains the constraints of considering possibilities of PC and Audio together: $O'_1 \cup O'_2=\{PC, A\}$.  $O_3  = O_1 \cup O'_1 \cup O_2 \cup O'_2$ contains the possibilities of PC, Audio and Video together: $O_3 = O_1 \cup O'_1 \cup O_2 \cup O'_2=\{PC, V, A\}$.  $O'_1$ contains the dimensions of PC that matter for connecting it with audio devices.
  $O'_2$ contains just the single object 'audio'.  Taken together, the structure underlying the digital hub is depicted in Fig. \ref{strucofhub}.\\

Functors, that is, structure preserving mappings allow us to express that
  constraints and possibilities that arise by linking observables in one
  domain apply to another domain as well. In this sense, structure
  preserving mapping express analogies between different areas of an
  overall strategy.

  Consider the example of an analogy between the PC-Video and the PC-Audio spoke of the digital hub. By identifying features and specific values in the presheaf over the PC-Audio observables with features and specific values in the presheaf over the PC-Video observables, we can see that a pattern of constraints that characterizes the iMovie section also characterizes the iTunes section. Formally, this is the construction of Lemma \ref{prefun}.\\
  Visually, this can be read from the upper part of  Fig. \ref{iTunes}, by comparison with figure \ref{imovie}.

\begin{figure}[h]\label{iTunes}
  \centering
  \includegraphics[scale=.4]{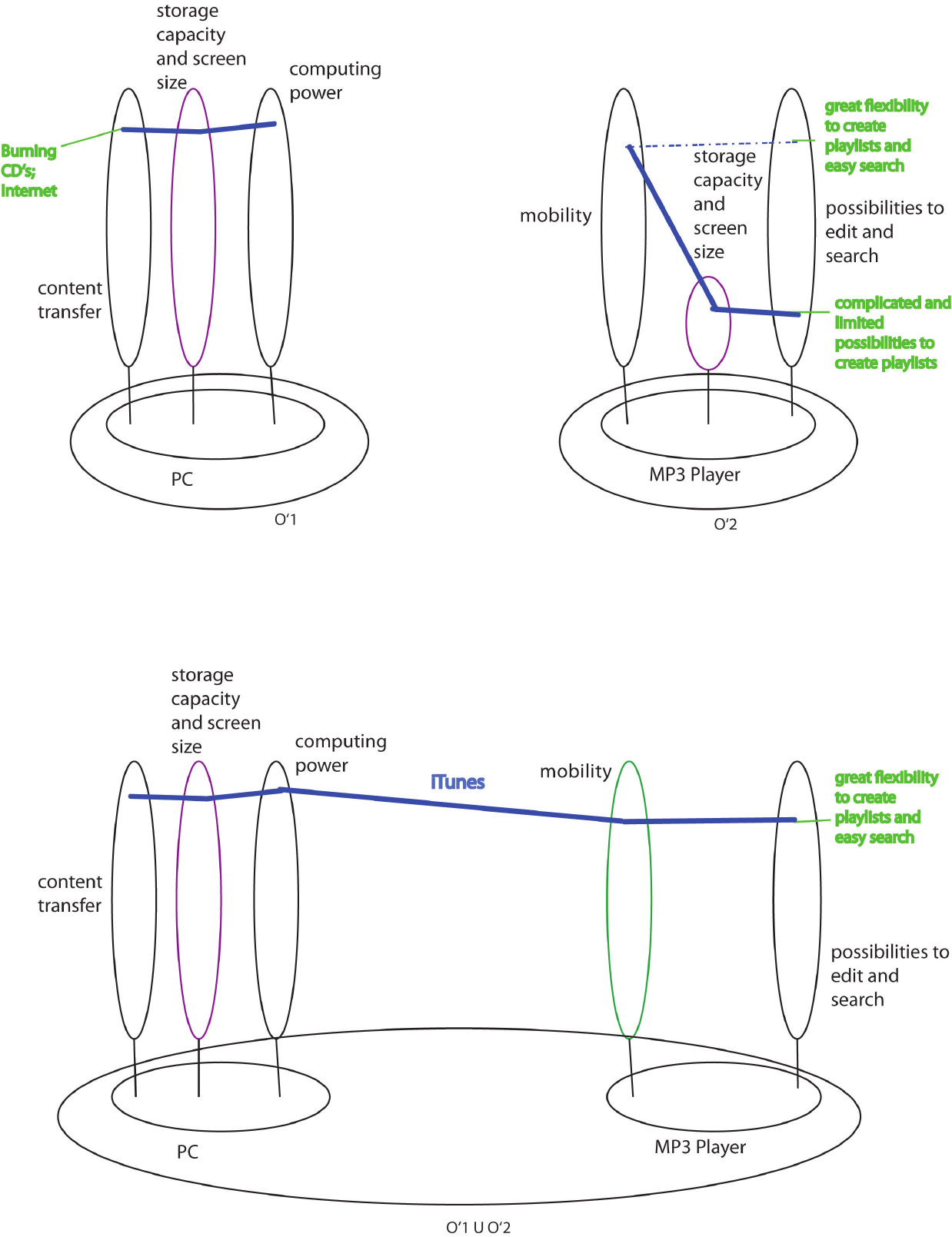}
  \caption{iMovie: a pictorial representation of the amalgamation of PC and
    MP3 player}
  \label{iTunes}
\end{figure}

Again, we can perform a merge operation as in Def. \ref{premerge}.
As indicated in figure \ref{iTunes}, there is one section contained in the presheaf over $Q(O'_1 \cup O'_2)$ (we again omit the original sections in $O'_1$ and $O'_2$ as they are no longer of interest):
\begin{eqnarray}
\label{itunessection}\nonumber
  &Q(O'_1 \cup O'_2) \supset &\\
\nonumber  \{ 
&\left( \begin{array}{c}
\text{music usage everywhere}\\
\text{storage capacity (harddisc) and Internet speed is large}\\
\text{computing power is large}
\end{array}
\right)\ ,&   \\
&\left(
\begin{array}{c}
\text{music usage everywhere}\\
\text{storage capacity is small}\\
\text{music bought and shared over the Internet}
\end{array}  \nonumber
\right)&  \}
\end{eqnarray}

By means of identifying observables in $O_1 \cup O_2$ with observables in
$O'_1 \cup O'_2$, and values of observables in $O_1 \cup O_2$ with values of
observables in $O'_1 \cup O'_2$, we can express that patterns of constraints
among observables in $O_1 \cup O_2$ match with patterns of constraints in
$O'_1 \cup O'_2$. In Fig. \ref{identify} we indicated which observables
in $PC-iTunes$ are similar to observables in $PC-iMovie$.  Moreover, we
identify values of observables in   $PC-iTunes$ with values of observables in
$PC-iMovie$.\\
 
\begin{figure}[h]
\centering
\caption{\textbf{Identification of similarities between $PC-iTunes$ and $PC-iMovie$}}
\begin{boxedminipage}{\textwidth}
\begin{center}
\textbf{Identifications observables in $PC-iTunes$  similar to observables in $PC-iMovie$:}\\
music content (in PC in $O'1 \cup O'2$) $\rightarrow$ film content (in PC in $O_1 \cup O_2$)\\
storage capacity (in PC in $O'1 \cup O'2$) $\rightarrow$ screen size (in PC in $O_1 \cup O_2$)\\
computing power (in PC in $O'1 \cup O'2$) $\rightarrow$ computing power (in PC in $O_1 \cup O_2$)\\
storage capacity (in MP3-player in $O'1 \cup O'2$)  $\rightarrow$ screen size (in Camcorder in $O_1 \cup O_2$)\\
possibilities to buy and share music (in MP3-player in $O'1 \cup O'2$) $\rightarrow$ possibilities to edit content (in Camcorder in $O_1 \cup O_2$)\\

\textbf{Identification of values in observables in $PC-iTunes$  similar to values of observables in $PC-iMovie$:}\\
'music usage everywhere' (value of 'music content' in 'MP3 Player')  $\rightarrow$ 'amateur and professional content' (value of 'film content' in camcorder) \\
'music usage everywhere'  (value of music content in PC in $O'1 \cup O'2$ )  $\rightarrow$ 'amateur and professional editing' (value of film content in PC in $O_1 \cup O_2$ ) \\
'music bought and shared over the Internet' (value of 'possibilities to share and buy music' in MP3-Player) $\rightarrow$ 'quick and easy editing' (value of 'possibilities to edit content' in Camcorder)\\
'only illegal music is shared' (value of possibility to buy and share music in MP3-Player) $\rightarrow$ 'difficult and inconvenient editing' (value of possibilities to edit content in Camcorder)\\

\end{center}

\end{boxedminipage}
\label{identify}
\end{figure}

 \begin{equation}\label{expand}
\begindc{\commdiag}[20]
\obj(70,40)[2]{$O_2$}
\obj(30,40)[1]{$O_1$}
  \obj(30,10)[3]{$QO_1$}
  \obj(70,10)[4]{$QO_2$}
  \obj(50,0)[5]{$O_3$}
    \obj(50,-20)[6]{$QO_3$}
  \mor{1}{2}{$f_{12}$}
  \mor{1}{3}{$g_1$}[\atright,\solidarrow]
   \mor{4}{3}{${f_{12}}^\ast$}[\atright,\solidarrow]
   \mor{2}{4}{$g_2$}
   \mor{2}{5}{$f_{23}$}[\atright,\solidarrow]
   \mor{1}{5}{$f_{13}$}
   \mor{6}{3}{$f_{13}^\ast$}
   \mor{6}{4}{$f_{23}^\ast$}[\atright,\solidarrow]
     \mor{5}{6}{$g_3$}
  \enddc
\end{equation} 
In this diagram, $f_{13}, f_{23}$ are simply subset relations, whereas $f_{12}$ indicates the functor described above. The functors between the presheaves come from Lemma \ref{prefun}.\\


Amalgamations in the iMovie spoke of the digital hub relate to amalgamations in the iTunes spoke of the digital hub. Above, we elaborated on the constraints in the iMovie spoke. For instance, it is a constraint for the local section in Camcorder in $O_2$ that 'possibilities to edit content' can only be filled with 'difficult and inconvenient editing', as the dimension 'screen size' takes the value 'small', and this constrains the value of 'possibilities to edit content'. By analogy, it is a constraint for the local section in MP3-Player in $O'_2$  that 'possibilities to buy and share music' can only be filled with 'only illegal music sharing', if the observable 'storage capacity and Internet access' takes the value 'small'. However, if the constraint imposed by 'small storage capacity and Internet access' is removed, the combination of the values 'music usage everywhere' in music content and 'music bought and shared over the Internet' in possibilities to buy and share music is admissible. That is expressed by the local section over 'music content' and 'possibilities to buy and share music' (blue dotted line), where the latter can take the value 'music bought and shared over the Internet'. Thus, this possibility is in principle available, but the further dimension 'storage capacity and Internet access' constrains it.\\

Now we come to an important property of the commuting diagrams we discussed above, which is in fact the defining property of analogies. The whole diagram (compare diagram \ref{appleanalogy} below) expresses that there is a structural similarity between the constraints in a source domain (which here is  the domain $O_1 \cup O_2$) and a target domain (which here is $O'_1 \cup O'_2$): the patterns of constraints are similar.  Just as the section in $O_1 \cup O_2$ in figure \ref{imovie} and the section in $O'_1 \cup O'_2$ in \ref{iTunes} are similar, the structural constraints between different features in $F O_1 \cup O_2$ and $F O'_1 \cup O'_2$ are similar. 
One could derive information about a constraint in $O'_1 \cup O'_2$ by looking at an identified constraint in $O_1 \cup O_2$. Once observables and values in the source and target domain are identified (we did this in figure \ref{identify}), the property of a commuting diagram tells us that we can derive information about the constraints among values in the target domain by looking at the source domain. 

By identifying features and specific values in the presheaf over the PC-Audio observables with features and specific values in the presheaf over the PC-Video observables, we can see that a pattern of constraints that characterizes the iMovie section also characterizes the iTunes section.
The analogy between the PC-Video spoke of the digital hub and the PC-Audio spoke of the digital hub is formally expressed in the following commuting diagram (see Lemma \ref{prefun}; the presheaf $Q'$ is determined by $Q$ and $h$):
\bel{appleanalogy}
\begin{CD}
    O'_1 \cup O'_2 @>h>> O_1 \cup O_2     \\
    @V{{Q'}}VV    @VV{Q}V     \\
    Q'(O'_1 \cup O'_2 )@<h^\ast<< Q (O_1 \cup O_2)\ ,
  \end{CD}
\qe 


For instance, if there is a section in $O_1 \cup O_2$, as expressed in equation \ref{imoviesection} above, the lower horizontal arrow expresses that there must be a corresponding section in $O'_1 \cup O'_2$. Indeed, the corresponding section exists and is expressed above in equation \eqref{itunessection}. 

The properties expressed by commuting diagrams run even deeper. As we map from one presheaf to another, we implicitly describe that local sections in the target domain also agree with local sections in the source domain. As discussed above, this is the case in our example, as the two amalgamations are similar. In sum, commuting diagrams express that two areas of a strategy share the deep structure of constraints. A target domain is set into relation with a source domain by the identification of similar observables (the upper arrow in \ref{appleanalogy}), with the aim of transferring structure of the source domain to the target domain (the lower arrow in \ref{appleanalogy}).

In order to identify such a commuting diagram, one needs to identify observables in two domains that correspond to it each other and possess the same relational structure, as in Fig. \ref{identify}. The commuting diagram thus formalizes the correspondence between the observables and their relations and represents them in a transparent manner.


\subsection{Transferring structure and preserving constraints}

If constructive analogy drives the expansion of the whole "digital hub" structure, one should be aware of the fact that by increasing the number of amalgams that are merged in the digital hub, one creates a richer web of constraints.

This implication can be again expressed more formally by diagrammatic
reasoning (see the diagram  \eqref{expand}).

$ O_1 \cup O_2$ is the first amalgam (PC and camcorder). $ O'_1 \cup O'_2$  is
the second one constructed by analogy (the PC-audio amalgam). $O_3$ is the
union of both amalgams, $ O_1 \cup O_2$ and  $ O'_1 \cup O'_2$.  $ O'_1 \cup
O'_2$ is constructed on the  $ O_1 \cup O_2$ template (f1). $O_3$ (a simple
hub-with two-spokes) is the union of $ O_1 \cup O_2$ and  $ O'_1 \cup O'_2$,
thus the arrows from $ O_1 \cup O_2$ to $O_3$ and $ O'_1 \cup O'_2$ to $O_3$
represent the inclusion relation. The arrow $g_1:  O_1 \cup O_2\rightarrow Q(O_1 \cup O_2)$  is the first amalgam presheaf, and so for $g_2$ and $g_3$. Formally, this can be described by natural transformations, see Def. \ref{def:Category5}, that is, morphisms between functors.

Notice the inversion of the direction of arrows $h_1$ and $h_2$ with regards to the direction of the inclusion ones. This inversion of direction simply represents the firing of constraints from the overall structure on its components. While the analogy supported by f1 and the union of the two amalgams are actively constructed by the decision maker, those constraints fire automatically. Once $ O_1 \cup O_2$  and $ O'_1 \cup O'_2$  are merged into $O_3$, the extension of the set of features is reflected in increased constraints over sections - a basic property of presheaves. In other words, once the PC-camcorder and the PC-audio amalgams are united, such union implies stronger constraints on the coherence of both (for example this implies that the operating system should be able to support simultaneously both types of amalgams, that interfaces should be more standardized etc).  Of course, the same logic applies as the embryonic "digital hub" of $O_3$ is further expanded to include new devices.

\section{Discussion}

Our first aim was to introduce a qualitative mathematical language to express
formally  innovation as a recombination of preexisting systems..   We have
introduced the notion of a presheaf together with its sections and operations
in the category of presheaves and have shown how they provide a potential for
a richer modeling of the structural processes underlying recombination.  Such formal tools also allow us to provide a simple and rigorous context within which defining the somehow elusive notion of coherence of a combination of features \cite{HS,Po2,Si1}.  Furthermore,  once a basic formal structure has been established, it has been natural to define on
its ground operations that can manipulate it, providing a sort of elementary
grammar of one of the fundamental creative processes of innovation.  Amalgamation of two structures modelled as presheaves provides a
simple and rigorous definition of a fundamental mechanism generating
novelty, by enabling sections that would be otherwise unfeasible if
the two source structures  were kept separated. In fact, this property emerges
directly from the definition of a presheaf.  Our structural
approach makes the generative mechanisms of recombination explicit.
To demonstrate the productivity of our approach we have applied it  to the
reconstruction of a specific case, by modelling the structure of a well-known
keynote address in which Steve Jobs sketched in 2001 the fundamental concept
of the digital hub.  The vision expressed by that concept drove in the following decade the Apple strategy and the emergence of new product categories.  Our example shows that it is possible to evolve the case study genre towards more formal, mathematically grounded  discourse. Of course, this is not substituting a more narrative approach, but provides strong complementarities with it and extends the possibilities of a  theoretical use of case studies.
Our  formal framework offers also a useful side result. It can enable clarifying the formal structure of analogy in innovation. In particular, the  tool of commutative diagrams and the associated notions of structure-preserving mappings, functors and natural transformations allow us  to understand in abstract terms how  analogy is based on structural similarity,  transfering the internal structure, and  not just the surface features, of  the source analogue\cite{GD,GLR}.  Furthermore, our notion of change operators  distinguishes the pure detection of similarity between existing objects from the constructive generation of similarity by using the source structure as a template to generate a new object.  Finally,  when the transferred  structure is characterized as a presheaf, we show how analogy generates a contravariant  direction of the structure mapping, that introduces new constraints over feasible sections, as the new structure has to be integrated with the source one. In more concrete words, the new possibilities conceived via analogy backfire as new constraints once the new business has to be integrated into the existing ones. While in this paper we have explicitly eschewed the structural aspects of
strategy representations, we believe that the formal concepts that we
introduced may provide significant building blocks to articulate cognitive
models of the ionovation process.  Such concepts make it possible to model
in general terms concepts capturing their internal structure
(and not just their dimensions). This could characterize  important aspects of
creative thinking. For example, the notion of a section  expresses the correlation structure of features, underlying many inferential processes in creative innovation.  Furthermore  we expect that the change operations we sketched may be fundamental building blocks for process models of  innovation, offering the fundamental components that are combined in the search for new concepts and the discovery of new opportunities. For example, the process of  generating the representation of new opportunities by amalgamation of pre-existing concepts  may be actually drawn from a combination of the  extension and the stretching operations.  
One might expect that the constraints arising form the use of constructive analogy might drive the actual search processes for coherent solutions  in implementing the original analogical insight into a viable configuration of features. 

Mathematics is not just about calculation and proofs, it is also about recognizing common patterns in different objects, and extracting their underlying formal structure. We hope that such an approach may help shaping more rigorous concepts in domains traditionally resistant to formalization, and may contribute to broaden the formal toolkit of the study of innovation.

\section{Acknowledgements}
 This paper draws on former discussions and joint work with many colleagues, for which we want to express our recognition and gratitude.  Peter Gärdenfors contributed to our thinking about conceptual combination. Timo Ehrig contributed with many conversations, that took shape also in a former joint working paper on a different although related subject \cite{EJW}]. Burkhard Schipper provided constructive criticism on an earlier draft.

\end{document}